\documentclass[rnote]{aa}
\usepackage[varg]{txfonts}

\usepackage{graphicx}
\usepackage{natbib}
\usepackage{color}

\begin{document}
   \title{The case for high precision in elemental abundances of stars\\ in the era of large spectroscopic surveys}
\titlerunning{The case for high precision in elemental abundances}


   \author{L. Lindegren
          \and
          S. Feltzing
          }

   \institute{Lund Observatory, Department of Astronomy and 
Theoretical Physics, Lund University, Box 43, SE-221\,00 Lund, Sweden\\
              \email{lennart \& sofia@astro.lu.se}
             }

   \date{Received 8 January 2013 / Accepted 11 April 2013}

 
  \abstract
  {A number of large spectroscopic surveys of stars in the Milky Way
    are under way or are being planned. In this context it is
    important to discuss the extent to which 
    elemental abundances can be used as discriminators between different (known
    and unknown) stellar populations in the Milky Way.}
   {We aim to establish the requirements in terms of precision in elemental
abundances, as derived from spectroscopic surveys
of the Milky Way's stellar populations, in order to detect
interesting substructures in elemental abundance space.}
   {We used Monte Carlo simulations to examine under which conditions 
   substructures in elemental abundance space can realistically be detected.}
   {We present a simple relation between the minimum number of stars needed 
   to detect a given substructure and the precision
   of the measurements. The results are in agreement
with recent small- and large-scale studies, with high and low precision, respectively.}
   {Large-number statistics cannot fully compensate for low precision in the
   abundance measurements. Each survey should carefully evaluate what the 
   main science drivers are for the survey and ensure that the chosen observational 
   strategy will result in the precision necessary to answer the questions posed.}


   \maketitle
%

\section{Introduction}

In Galactic archaeology stars are used as time capsules. The outer layers of 
their atmospheres, accessible to spectroscopic studies, are presumed 
to keep a fair representation of the mixture of elements present in the gas 
cloud out of which they formed. By
determining the elemental abundances in the stellar photospheres, and
combining with kinematics and age information, it is possible to
piece together the history of the Milky Way
\citep[e.g.,][]{freeman2002}.

\citet{edvardsson1993} were among the first to
demonstrate that it is possible to achieve high precision in studies
of elemental abundances for large samples of long-lived dwarf stars. In
their study of 189 nearby dwarf stars they achieved a precision%
\footnote{The term `precision' refers to the ability of a method to give
the same result from repeated measurements (observations). See Sect.~\ref{sect:ap} and 
footnote~\ref{ftn:acc} for the distinction between precision and accuracy.} 
better
than 0.05~dex for many elements, including iron and nickel. More recent
studies have achieved similar or even better
results. \citet{nissen2010}, in a study of dwarf stars in the halo,
obtain a precision of 0.02--0.04~dex for $\alpha$-elements relative to
iron, and 0.01~dex for nickel relative to iron. In studies of solar
twins (i.e., stars whose stellar parameters, including metallicity, closely
match those of the sun) \citet{melendez2012} are able to
achieve a precision better than 0.01~dex. At the same
time several studies have found that in the solar neighbourhood there
exist substructures in the elemental abundance trends with differences
as large as 0.1 to 0.2~dex
\citep[e.g.][]{fuhrmann1998,bensby2004,nissen2010}.

Driven both by technological advances and the need for ground-based
observations to complement and follow up the expected observations
from the Gaia satellite, Galactic astronomy is entering a new regime
where elemental abundances are derived for very large samples of
stars. Dedicated survey telescopes and large surveys using existing
telescopes have already moved Galactic astronomy into the era
of large spectroscopic surveys
\citep{RAVE2,SEGUE,2010IAUS..265..480M,gilmore2012}.  With the new
surveys, several hundred thousands of stars will be observed for each
stellar component of the Galaxy. For all of these stars we will have
elemental abundances as well as kinematics and, when feasible, ages.
One goal for these studies is to quantify the extent to which the
differences in elemental abundances seen in the solar neighbourhood
extend to other parts of the stellar disk(s) and halo, and to identify
other (as yet unknown) components that may exist here and elsewhere.

Large-scale surveys naturally tend to have lower signal-to-noise
ratios for the individual stars than can be achieved in the classical
studies of small stellar samples in the solar neighbourhood. On the
other hand, the very large number of stars reached with the new
surveys will at least partly compensate for a lower precision per
object.  A relevant question is thus: How many stars do we need to
detect a certain abundance signature of $Y$~dex, when we have a
precision of $Z$~dex in the individual abundance determinations?
This is what we explore in this \textit{Research Note}.

This Research Note is structured as follows: Section~\ref{sect:def} sets out
the problem which is then investigated in Sect.~\ref{sect:invest}. In
Sect.~\ref{sect:ap} we discuss what accuracies and precisions have
been shown to be possible and what is feasible to expect from large
scale surveys. Section~\ref{sect:concl} contains some concluding
remarks.

%

\section{Defining the problem}
\label{sect:def}

\begin{figure}
\resizebox{\hsize}{!}{\includegraphics[]{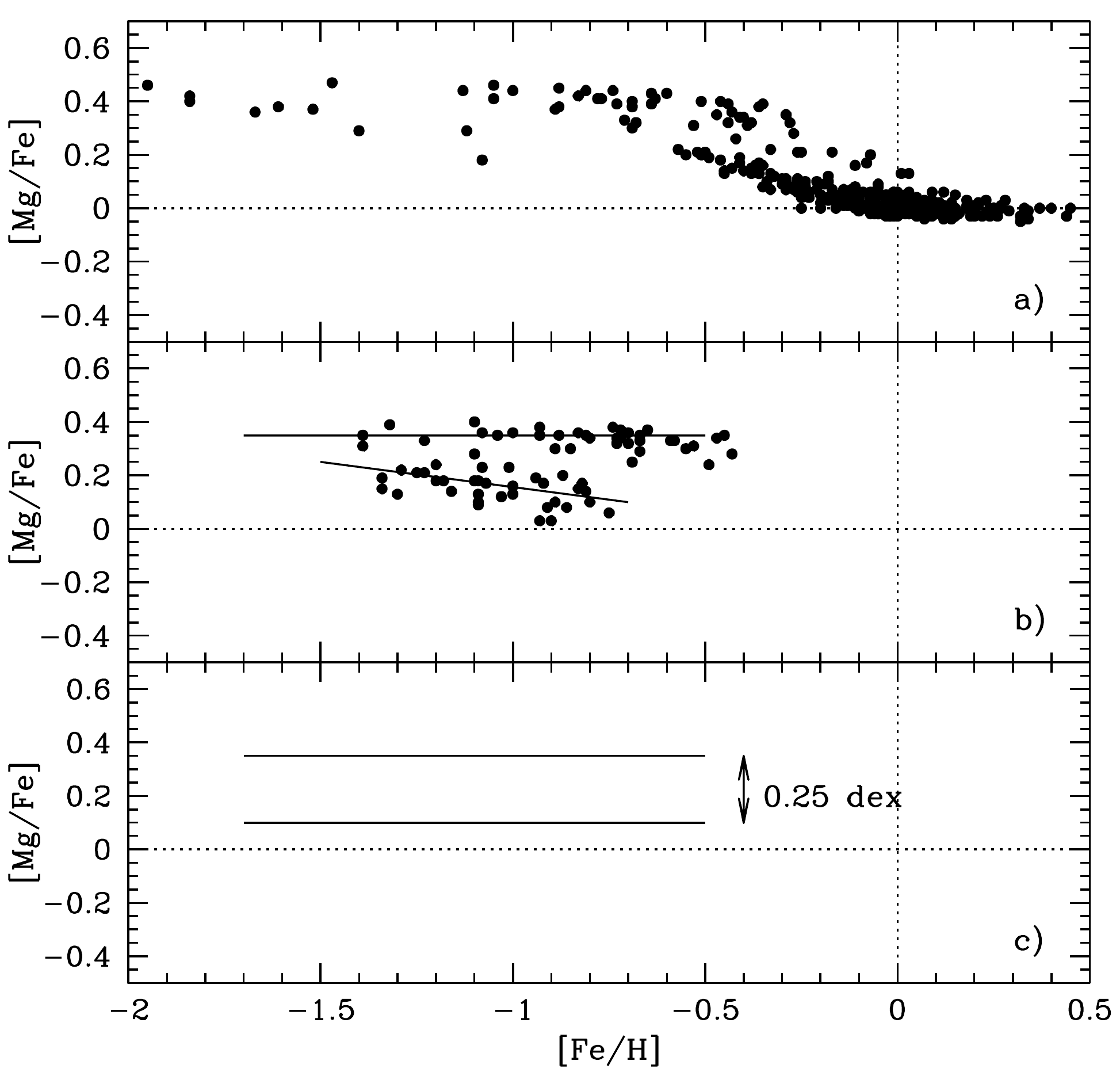}}
\caption{Illustration of the problem, showing Fe and Mg abundances for
  stars in the solar neighbourhood. {\bf a} Based on data by Fuhrmann
  (see text for references). At each value of [Fe/H] the stars fall into two groups with distinctly different
[Mg/Fe]. {\bf b} Based on data for stars with halo velocities from \citet{nissen2010}.
The two lines, drawn by hand, illustrate the separation in high- and low-$\alpha$
stars identified by \citet{nissen2010}.
{\bf c} Illustration of the generic problem treated here.}
\label{fig:ill1}
\end{figure}

Elemental abundances derived from stellar spectra with high resolution
and high signal-to-noise ratios have shown that the stars in the Milky
Way and in the nearby dwarf spheroidal galaxies have a range of
elemental abundances \citep[see, e.g.,][]{2009ARA&A..47..371T}. Not
only do the stars span many orders of magnitude in iron abundances
([Fe/H]\footnote{We use the standard notation for elemental abundances
where [Fe/H] $=\log{(N_{\rm Fe}/N_{\rm H}})_{\ast} - \log{(N_{\rm
Fe}/N_{\rm H}})_{\odot}$.}) they also show, subtler, differences
in relative abundance. One of the most well-known examples is given by
the solar neighbourhood, where for example \citet{fuhrmann1998,fuhrmann2000,fuhrmann2002,fuhrmann2004,fuhrmann2008,fuhrmann2011} shows
from a basically volume limited sample that there are two abundance
trends present. One trend has high [Mg/Fe] and one with low, almost
solar, [Mg/H]. Figure~\ref{fig:ill1}a reproduces his results.
The basic result, i.e., that there is a split in the abundance trends
was foreshadowed by several studies \citep[e.g.,][]{edvardsson1993}
and has been reproduced by a number of studies since
\citep[e.g.,][]{reddy2003,bensby2004,bensby2005,reddy2006,neves2009,2012A&A...545A..32A}.
Another well-known example in the solar neighbourhood is the split in
$\alpha$-elements as well as in Na and Ni for stars with typical halo
kinematics \citep[][and Fig.~\ref{fig:ill1}b]{nissen2010}.
The differences in elemental abundances between these different
populations can be as large as 0.2~dex, but often they are smaller.

Figure~\ref{fig:ill1}c illustrates the highly simplified case
considered in the present study, namely that the observed stars belong
to two populations that differ in some abundance ratio
[X/Fe] by a certain amount. In the figure the difference is taken to
be 0.25~dex, which as we have seen may be representative of actual
abundance differences. We will investigate whether it is possible to
distinguish the two populations depending on the number of stars
considered and the precision of the individual [X/Fe] measurements.
This will allow us to derive a lower limit for the precision needed
to probe abundance trends such as those shown in
Fig.~\ref{fig:ill1}. We emphasize that the objective is to 
identify such substructures in elemental abundance space without
a~priori categorization of the stars, e.g., in terms of kinematic
populations.

%

\section{Investigation}
\label{sect:invest}

The problem is formulated as a classical hypothesis test. Although 
hypothesis testing is a well-known technique, and the present 
application follows standard methodology, we describe our 
assumptions and calculations in some detail in order to provide a good
theoretical framework for the subsequent discussion.  

Consider a sample of $N$ stars for which measurements 
$x_i$, $i=1,\dots,N$ of some quantity $X$ (e.g., [Mg/Fe]) have been made 
with uniform precision. The null hypothesis $H_0$ is that there is just a single 
population with fixed but unknown mean abundance $\mu$ (but possibly with 
some intrinsic scatter, assumed to be Gaussian). Assuming that the 
measurement errors are unbiased and Gaussian, the values $x_i$ are thus 
expected to scatter around $\mu$ with some dispersion $\sigma$ which
is essentially unknown because it includes the internal scatter as well as the
measurement errors. The alternative hypothesis $H_A$ is that the stars are 
drawn from two distinct and equally large populations, with mean values 
$\mu_1$ and $\mu_2$, respectively, but otherwise similar properties. In particular,
the intrinsic scatter in each population is the same as in $H_0$, and the
measurement errors are also the same. Without loss of generality we may
take $\mu=0$ in $H_0$, and $\mu_{1,2}=\pm r\sigma/2$ in $H_A$, so 
that the populations are separated by $r>0$ standard deviations in $H_A$, 
and by $r=0$ in $H_0$. The only relevant quantities to consider are
then the (dimensionless) separation $r\ge 0$ and the total size of the sample 
$N$. 

The possibility to distinguish the two populations in $H_A$ depends both 
on $r$ and $N$. Clearly, if $r$ is large (say $>5$) the two populations will
show up as distinct even for small samples (say $N = 100$ stars). For
smaller $r$ it may still be possible to distinguish the populations if $N$ 
is large enough. Exactly how large $N$ must be for a given $r$ is what we 
want to find out. Conversely, for a given $N$ this will also show the 
minimum $r$ that can be distinguished. Given the true separation in
logarithmic abundance (dex), this in turn sets an upper limit on the 
standard error of the abundance measurements.

\begin{figure}
\resizebox{\hsize}{!}{\includegraphics[]{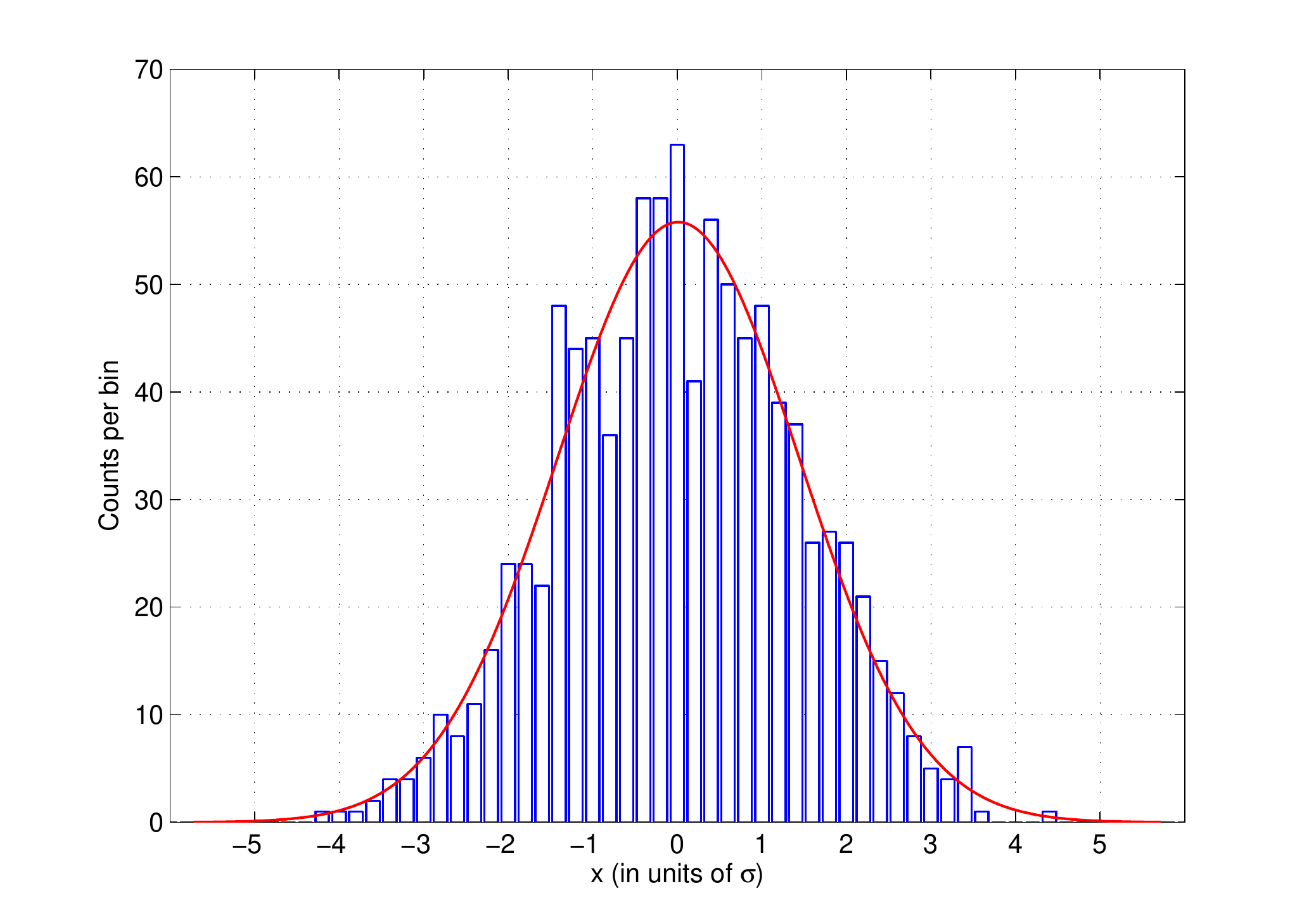}}
\resizebox{\hsize}{!}{\includegraphics[]{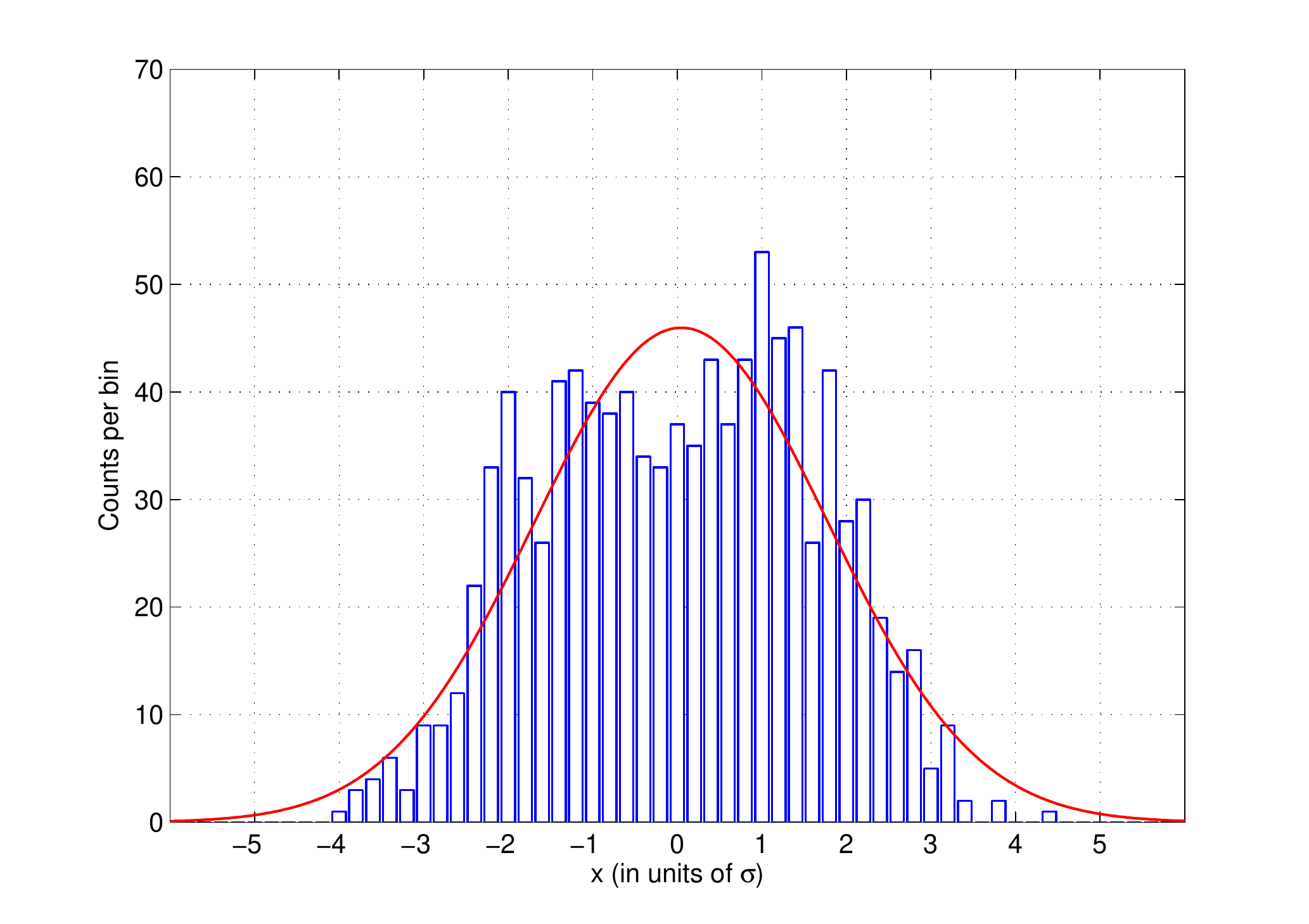}}
\caption{The top histogram shows a simulated sample of size $N=1000$
drawn from a superposition of two Gaussian distributions separated
by $r=2.0$ standard deviations. The solid curve is the best-fitting single
Gaussian. In this case the null hypothesis, that the sample was drawn from
a single Gaussian, cannot be rejected. The bottom histogram shows a
simulation with separation $r=2.4$ standard deviations. The solid curve is
again the best-fitting single Gaussian. In this case the
null hypothesis would be rejected and a much better fit could be obtained
by fitting two Gaussians (not shown).}
\label{fig:hist1000}
\end{figure}

The two simulated samples in Fig.~\ref{fig:hist1000} illustrate 
the situation for $N=1000$. In the top diagram (generated with $r=2.0$) 
it is not possible to conclude that there are two populations, while in the 
bottom one (for $r=2.4$) they are rather clearly distinguished.

Given the data $\vec{x}=(x_1,\,x_2,\,\dots,\,x_N)$ we now compute a
test statistic $t(\vec{x})$ quantifying how much the data deviate from the 
distribution assumed under the null hypothesis, i.e., in this case
a Gaussian with mean value $\mu$ and standard deviation $\sigma$
(both of which must be estimated from the data). A large value of $t$ 
indicates that the data do not follow this distribution. The null 
hypothesis is consequently rejected if $t(\vec{x})$
exceeds some critical value $C$, chosen such that the probability of falsely 
rejecting $H_0$ is some suitably small number, say $\alpha=0.01$ (the 
significance of the test). 
  
It should be noted that $H_0$ and $H_A$ are not complementary, i.e., if
$H_0$ is rejected it does not automatically follow that $H_A$ should be 
accepted. Indeed, there are obviously many possible distributions of $X$ 
that cannot be described by either $H_0$ or $H_A$. Having rejected $H_0$, 
the next logical step is to test whether $H_A$ provides a reasonable 
explanation of the data, or if that hypothesis, too, has to be rejected. 
However, since we are specifically interested in detecting substructures
in the distribution of $X$, of which $H_A$ provides the simplest possible
example, it is very relevant to examine how powerful the chosen test is
in rejecting $H_0$, when $H_A$ is true, as a function of $N$ and $r$.

The test statistic $t(\vec{x})$ measures the ``distance'' of the data from 
the best-fitting normal (Gaussian) distribution with free parameters 
$\mu$ and $\sigma$. Numerous tests for ``normality'' exist, but many of
them are quite sensitive to outliers (indeed, some are constructed to detect 
outliers) and therefore unsuitable for our application. Instead we make use 
of the distance measure $D$ from the well-known Kolmogorov--Smirnov 
(K--S) one-sample test \citep{book:nr3}, which is relatively insensitive to outliers
and readily adapted to non-Gaussian distributions, if needed. We define
\begin{equation}\label{eq:t}
t(\vec{x}) = \sqrt{N}\times \min_{\mu,\,\sigma} \, \max_{x} \, 
\bigl| F_N(x;\,\vec{x}) - F(x;\,\mu,\sigma) \bigr| \, ,
\end{equation}
where $F_N$ is the empirical distribution function for the given data 
(i.e., $F_N(x;\,\vec{x})=n(x)/N$, where $n(x)$ is the number
of data points $\le x$) and $F(x;\,\mu,\sigma)$ is the normal cumulative
distribution function for mean value $\mu$ and standard deviation $\sigma$.
The expression in Eq.~(\ref{eq:t}) requires some explanation. The quantity 
obtained as the maximum of the absolute difference between the two
cumulative distributions is the distance measure $D$ used in the standard
one-sample K--S test. This $D$ is however a function of the parameters of the  
theoretical distribution, in this case $\mu$ and $\sigma$, and we therefore
adjust these parameters to give the minimum $D$. This is multiplied by
$\sqrt{N}$ to make the distribution of $t$ under $H_0$ nearly independent
of $N$, and to avoid inconveniently small values of $D$ for large samples.

\begin{figure}
\resizebox{0.9\hsize}{!}{\includegraphics[]{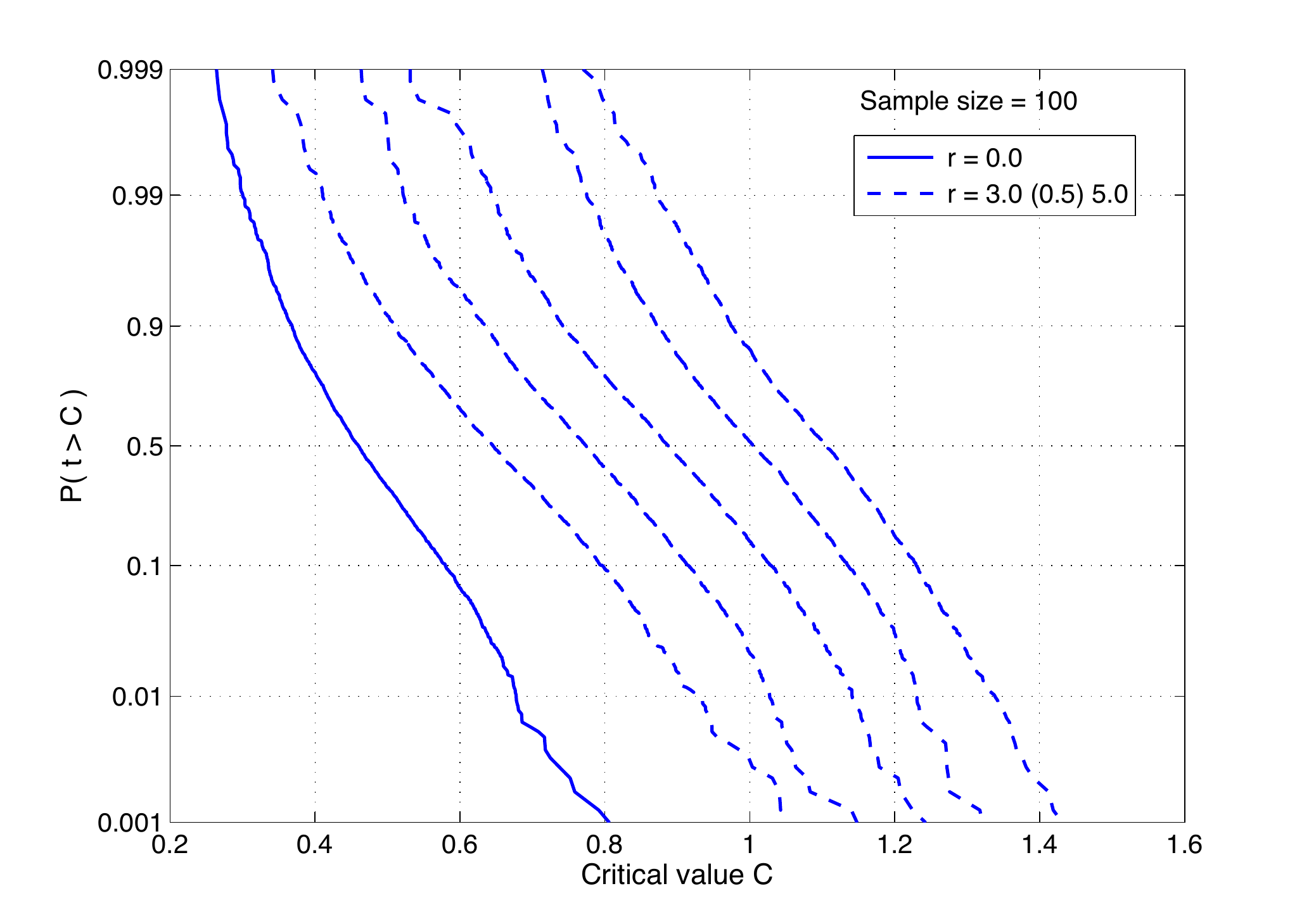}}
\resizebox{0.9\hsize}{!}{\includegraphics[]{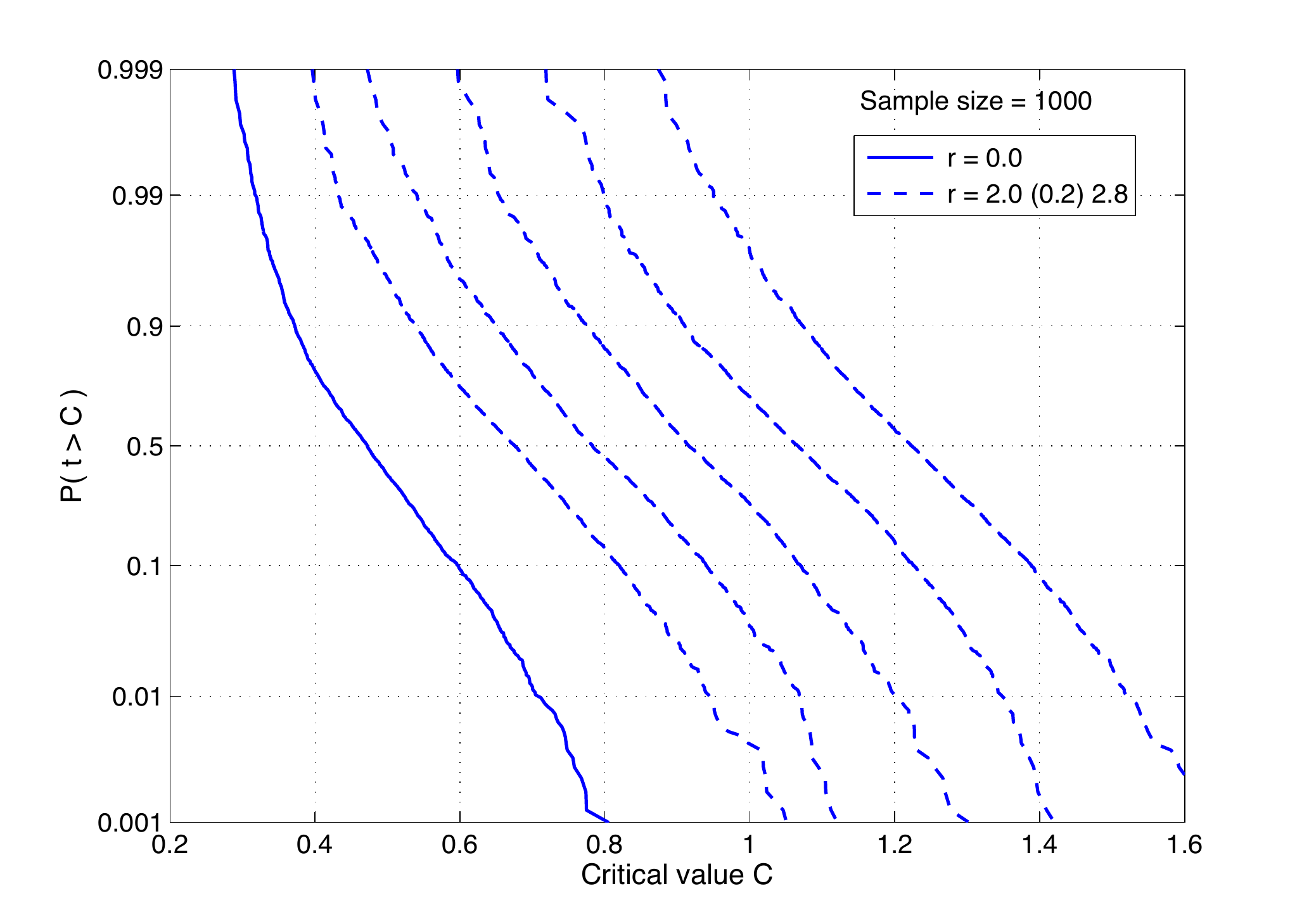}}
\resizebox{0.9\hsize}{!}{\includegraphics[]{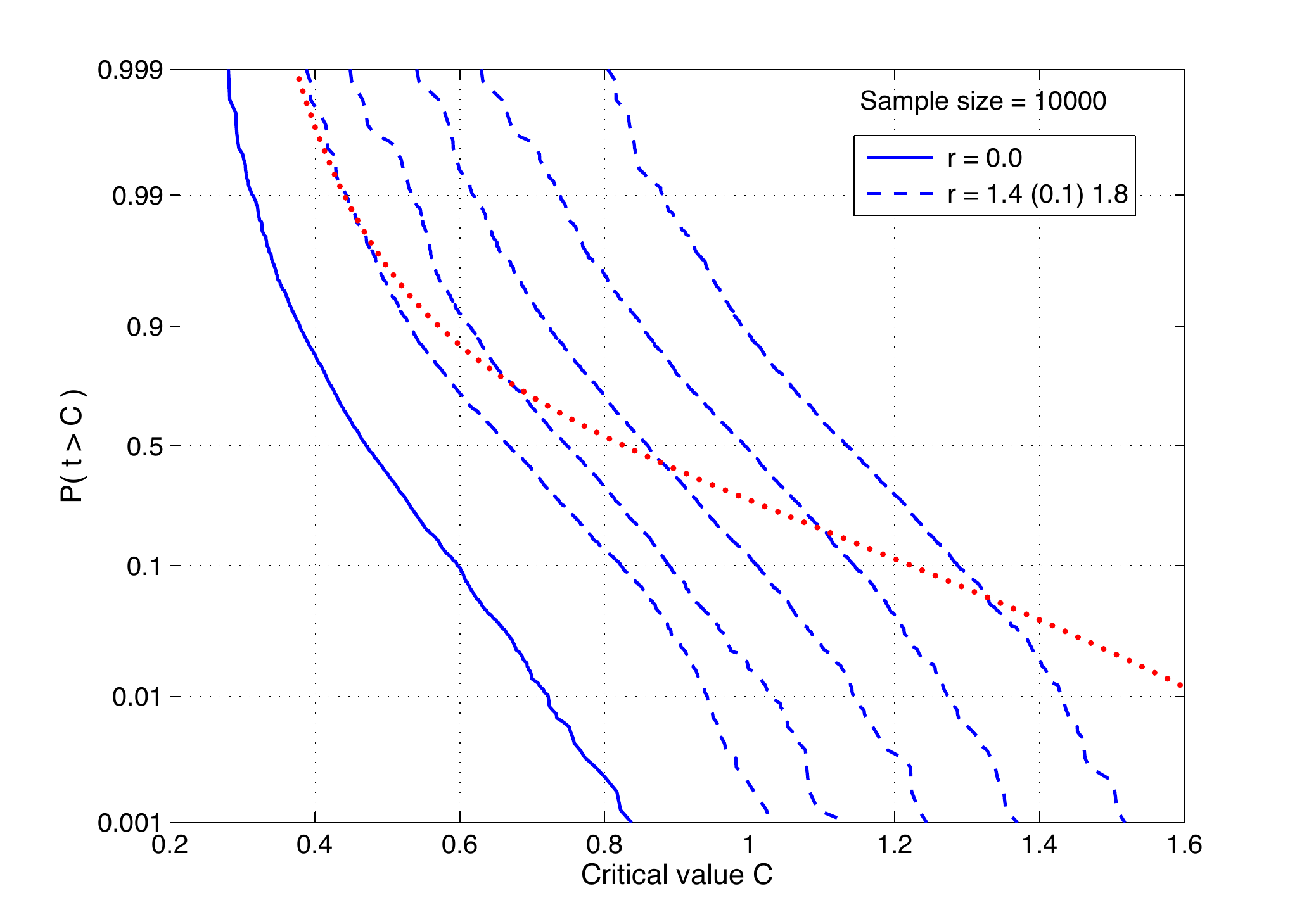}}
\caption{Examples of probability plots for the test statistic $t(\vec{x})$
obtained in Monte-Carlo simulations for sample sizes $N=10^2$,
$10^3$, and $10^4$ (top to bottom). In each diagram the solid curve
shows, as a function of the critical value $C$, the probability that $t$ 
exceeds $C$ under the null hypothesis ($r=0$). The dashed curves show 
the probabilities under the alternative hypothesis ($r>0$) for the 
$r$-values indicated in the legend. In the bottom diagram the dotted 
curve gives, for comparison, the expected distribution of $D\sqrt{N}$ for a 
one-sample K--S test in which $F$ is the true distribution (without adjusting
$\mu$ and $\sigma$); see footnote~\ref{footnote3}.}
\label{fig:distrT}
\end{figure}

\begin{figure}
\resizebox{0.9\hsize}{!}{\includegraphics[]{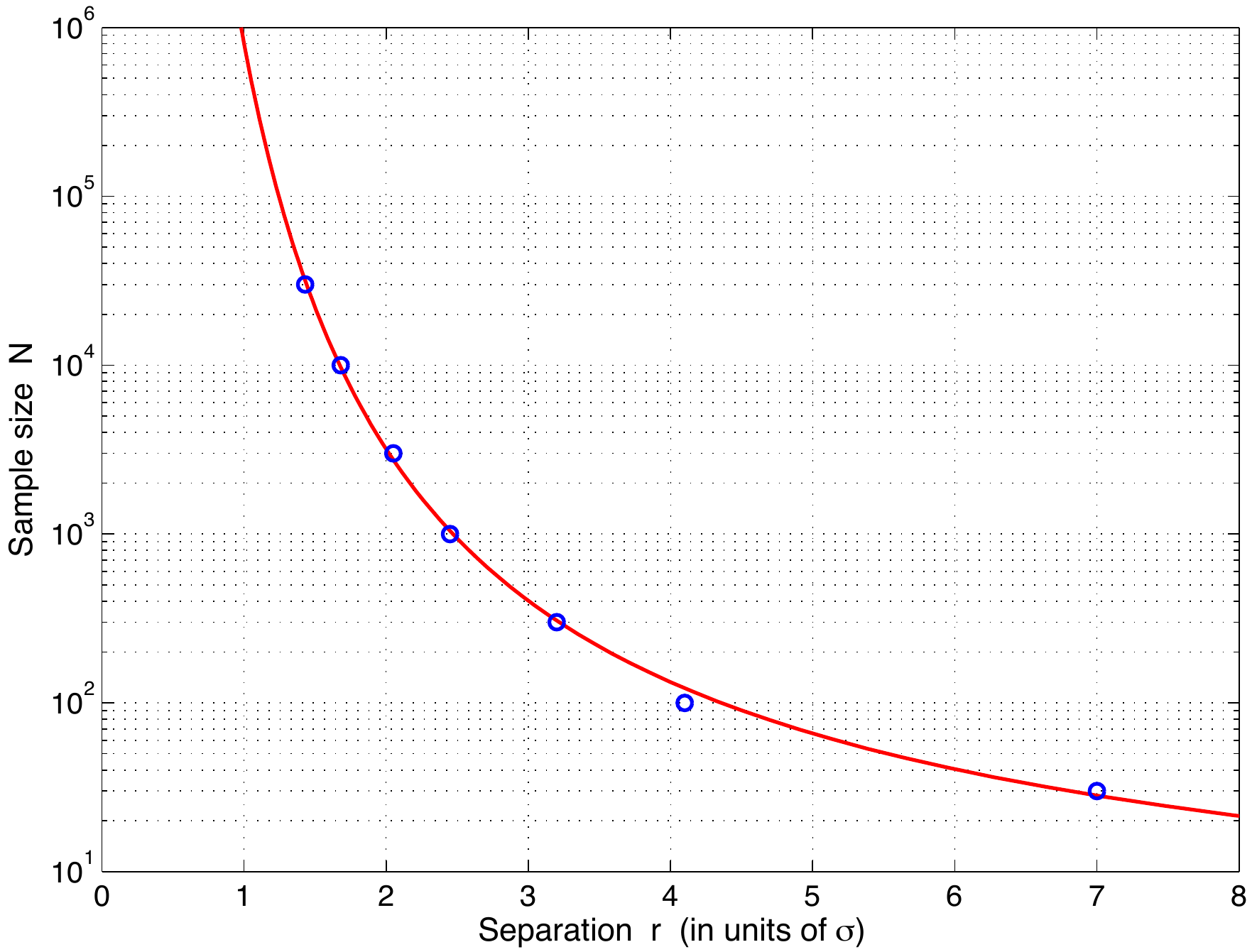}}
\caption{Minimum sample size needed to distinguish two equal
Gaussian populations, as a function of the separation of the population
mean in units of the standard deviation of each population. The circles
are the results from Monte-Carlo simulations as described in the text,
using a K--S type test with significance level $\alpha=0.01$ and
power $1-\beta=0.99$. The curve is the fitted function in Eq.~(\ref{eq:Nmin})
or (\ref{eq:Rmin}).}
\label{fig:NvsR}
\end{figure}

The distribution of $t(\vec{x})$ for given $N$ and $r$ must be determined 
through Monte-Carlo simulations, in which many independent realizations 
of $\vec{x}$ are generated and $t$ computed for each of them by application
of Eq.~(\ref{eq:t}).%
\footnote{The distribution of $t$ under $H_0$ does not follow
the theoretical distribution of $D\sqrt{N}$ usually given for the K--S test, 
i.e., $P(t>C)=Q_{KS}[(1+0.12/\sqrt{N}+0.11/N)C]$, where the function
$Q_{KS}$ is given in \citet{book:nr3}. This distribution, shown as a dotted 
curve in the bottom diagram of Fig.~\ref{fig:distrT}, is clearly very different from
the empirical distribution for $r=0$ given by the solid curve in the same diagram.
The reason is that the K--S test assumes that the comparison is 
made with a \emph{fixed} distribution $F(x)$. In our case we adjust $\mu$ and 
$\sigma$ to minimize $D\sqrt{N}$, which results in a different distribution.\label{footnote3}}
We give results for some selected combinations of $(N,r)$ in Fig.~\ref{fig:distrT}.
Each curve in these diagrams shows the fraction of $t$-values exceeding
$C$ in a simulation with 2000 realizations of $\vec{x}$. The fractions are 
plotted on a non-linear vertical scale (using a $\log[P/(1-P)]$ transformation) in order 
to highlight both tails of the distribution. The wiggles in the upper and lower
parts of the curves are caused by the number statistics due to the limited 
number of realizations.

For a given value of $C$, the significance of the test, i.e., the 
probability of falsely rejecting $H_0$ (``Type~I error''), can be directly read 
off the solid curves in Fig.~\ref{fig:distrT} as $\alpha=P(t>C;\,r=0)$. 
Conversely, we can determine the $C$-value to be used for a given significance 
level. Adopting a relatively conservative $\alpha = 0.01$ we find 
that $C\simeq 0.7$ can be used for any sample size.
For $r>0$ the dashed curves give the power $1-\beta$ of the test, where 
$\beta$ is the probability of a ``Type~II error'', i.e., of failing to reject $H_0$ 
when $H_A$ is true. For example, if we require $1-\beta\ge 0.99$ at 
$C=0.7$, the minimum $r$ that is detected with this high degree of probability 
is about 4.2, 2.5, and 1.7 for the sample sizes shown in Fig.~\ref{fig:distrT}.
For the two specific examples in Fig.~\ref{fig:hist1000} the computed statistic 
is $t=0.41$ (top) and 1.04 (bottom), meaning that $H_0$ would be rejected
at the 1\% significance level in the latter case, but not in the former.

Results are summarized in Fig.~\ref{fig:NvsR}, which shows the minimum
sample size as a function of $r$ for the assumptions described above. The
circles are the results of the Monte-Carlo simulations for
$\alpha=0$ and $1-\beta=0.99$, obtained by interpolating in 
Fig.~\ref{fig:distrT} and the corresponding diagrams for
$N=30$, 300, 3000, and 30\,000. The curve is the fitted function
\begin{equation}\label{eq:Nmin}
N_{\rm min} \simeq \exp\left( 0.6 + 13r^{-0.8} \right) \, .
\end{equation}
This function, which has no theoretical foundation and therefore should
not be used outside of the experimental range ($30 \le N \le 30\,000$), can be
inverted to give the minimum separation for a given sample size:
\begin{equation}\label{eq:Rmin}
r_{\rm min} \simeq \left( \frac{\ln N - 0.6}{13} \right) ^{-1.25} \, .
\end{equation}
For example, if the populations are separated by 5 times the measurement 
error ($r = 5$), the populations could be separated already for $N\simeq 70$. 
For $r = 3$ the minimum sample size is $N = 400$, and for $r = 2$ it is 
$N = 3000$.
Clearly, if the separation is about the same as the measurement errors
($r = 1$), the situation is virtually hopeless even if the sample includes
hundreds of thousands of stars.

It should be remembered that these results were obtained with a very specific
set of assumptions, including: 
(1) measurement errors (and/or internal scatter) that are purely Gaussian; 
(2) that the two populations in the alternative hypothesis are equally large;
(3) the use of the particular statistic in Eq.~(\ref{eq:t}); and 
(4) the choice of significance (a probability of falsely rejecting $H_0$ less than 
$\alpha=0.01$) and power (a probability of correctly rejecting $H_0$ greater than 
$1-\beta=0.99$).
Changing any of these assumptions would result in a different 
relation%
\footnote{Experiments with unequally large populations in $H_A$
suggest that the power of the test is not overly sensitive to this assumption, 
as long as there is a fair number of stars from each population in the sample.} 
 from the one shown in Fig.~\ref{fig:NvsR}. Nevertheless, this 
investigation already indicates how far we can go in replacing spectroscopic
resolution and signal-to-noise ratios (i.e., small measurement errors) with
large-number statistics. In particular when we consider that real data are 
never as clean, nor the expected abundance patterns as simple as assumed
here, our estimates must be regarded as lower bounds to what can
realistically be achieved.

%

\section{Accuracy and precision in stellar abundances}
\label{sect:ap}

\begin{figure}
\resizebox{0.85\hsize}{!}{\includegraphics[]{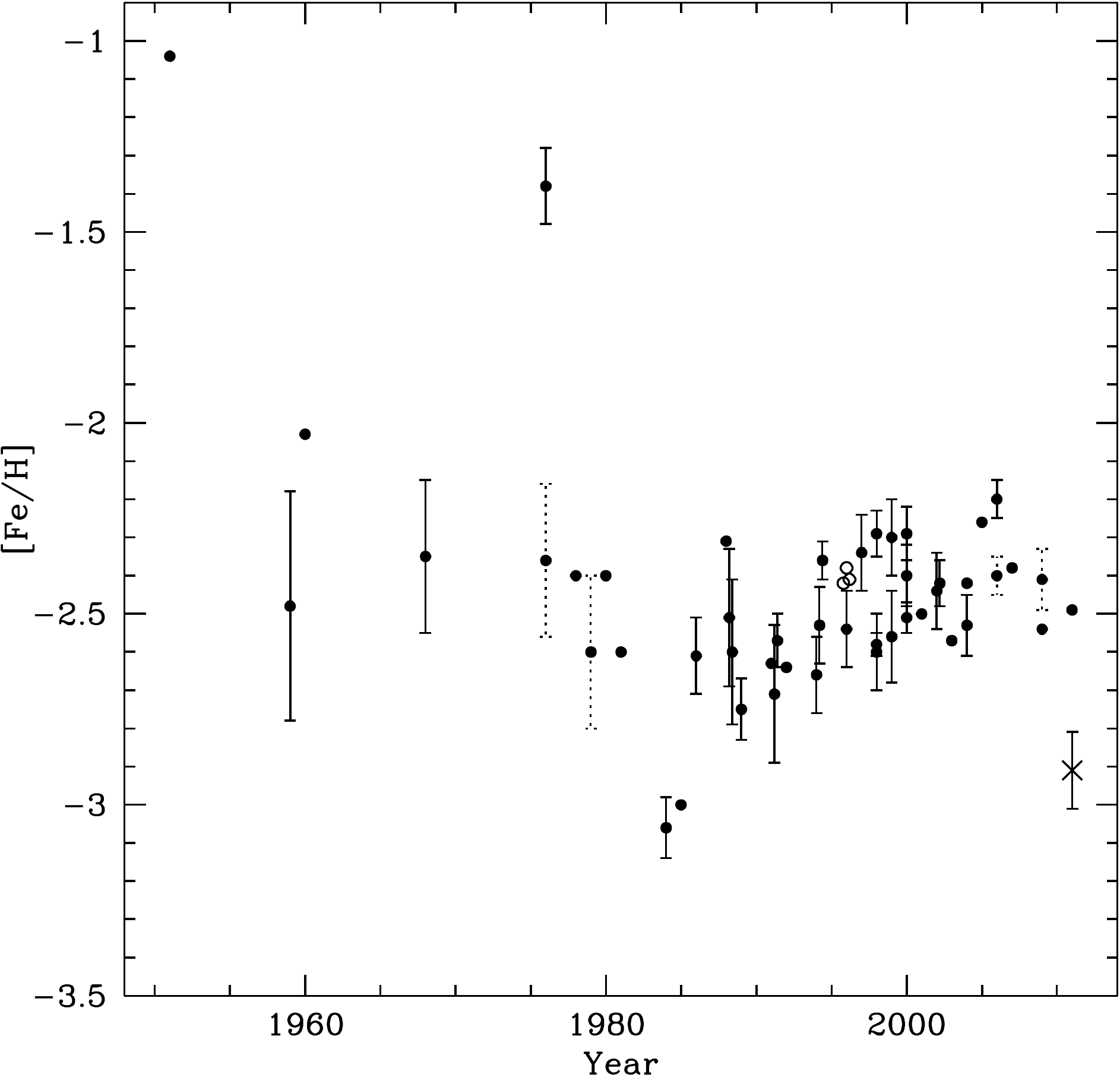}}
\caption{Updated plot for HD\,140283 from \citet{gustafsson2004}.  On
  the $y$ axis are shown the [Fe/H] values determined from spectroscopy
  as reported in the literature. The $x$ axis gives the year of the
  publication. The error bars indicate (full lines) the line-to-line
  scatter (when available) and (dotted lines) a few examples of
  attempts to address the full error including errors in the stellar
  parameters.  The circles ($\circ$, at 1996) refer to a single analysis using three
  different temperature scales and the cross ($\times$) refers to an analysis of a
  high-resolution spectrum using the SEGUE pipeline \citep[][see also
  Sect.~\ref{sect:ap} for database usage]{lee2011L}.  }
\label{fig:hd140283}
\end{figure}

We have no knowledge \textit{a priori} of the properties of a star and
no experiment to manipulate in the laboratory but can only observe the
emitted radiation and from that infer the stellar properties.
Therefore the accuracy%
\footnote{`Accuracy' refers to the capability of a method to return the 
correct result of a measurement, in contrast to precision which only
implies agreement between the results of different measurements. 
It is possible to have high precision but poor accuracy, as is often the 
case in astronomy. For the purpose
  of the study of trends in elemental abundances in the Milky Way both
  are important, but for practical reasons most studies are
  concerned with precision rather than accuracy.\label{ftn:acc}} 
of elemental
abundances in stars is often hard to ascertain as it depends on a
number of physical effects and properties that are not always
well-known, well-determined, or well-studied \citep[][]{baschek1991}.
Important examples of relevant effects include deviations from local
thermodynamic equilibrium (NLTE) and deviations from 1D geometry
\citep{asplund2005araa,2006A&A...452.1039H}. Additionally, systematic
and random errors in the stellar parameters will further decrease the
accuracy as well as the precision within a study.

\begin{figure}
\resizebox{0.85\hsize}{!}{\includegraphics[]{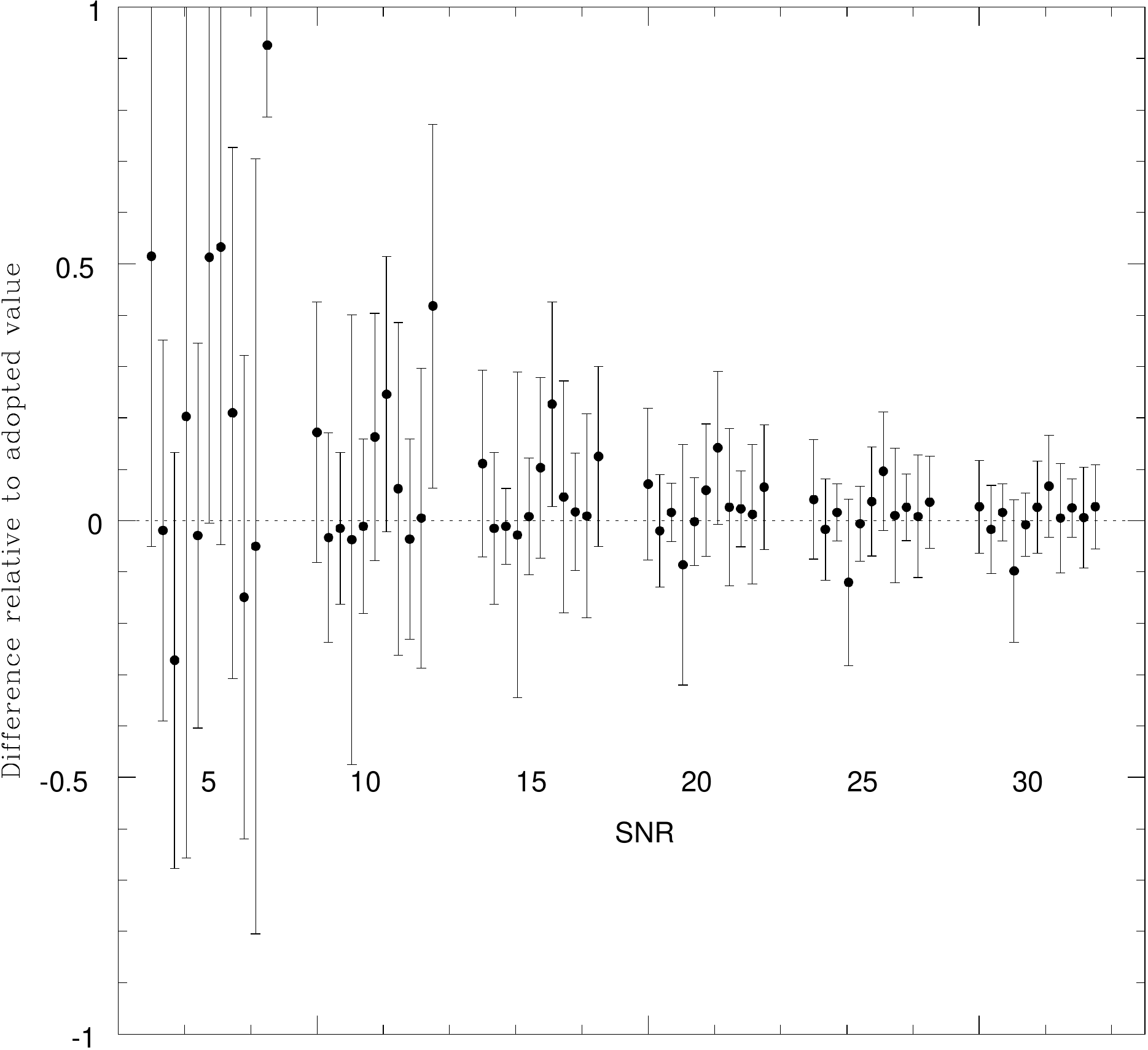}}
\caption{This figure shows results from \citet{lee2008}. We plot
 the difference between the final abundance adopted in \citet{lee2008}
and the abundance derived using one of eleven methods \citep[as described 
in][]{lee2008} as a function of the signal-to-noise ratio. The error bar 
shows the scatter for each method.}
\label{fig:lee2008}
\end{figure}

An interesting example of the slow convergence of the derived iron
abundance in spite of increasing precision is given in
\citet{gustafsson2004}, where he compares literature results for the
well studied metal-poor sub-giant HD\,142083. Over time the error-bars
resulting from line-to-line scatter decreases thanks to increased
wavelength coverage (i.e., more Fe\,{\sc i} lines are used in the
analysis) and higher signal-to-noise ratios. However, the differences
between studies remain large. An updated and augmented version of the
plot in \citet{gustafsson2004} is given in Fig.~\ref{fig:hd140283}.
Data were sourced using SIMBAD and the SAGA database
\citep{2008PASJ...60.1159S}.  For data listed in the SAGA database we
excluded all non-unique data, e.g., where a value for [Fe/H] is quoted
but that value is not determined in the study in question but taken
from a previous study.  The error bars shown are measures of the
precision based on the quality of the spectra and reflect the
line-to-line scatter.  Generally, the precision has clearly improved with 
time, but, judging from the scatter between different determinations, 
it is doubtful if the overall accuracy has improved much. From around 1995
most studies quote a precision from measurement errors and errors in
$\log gf$ values alone of 0.1~dex. From about the same time there appears also 
to be a convergence on two different [Fe/H] values. The difference is mainly
related to a high and a low value of $\log g$, whilst $T_{\rm eff}$
appears uncorrelated with this split in [Fe/H] values. This illustrates
the need for homogeneous samples treated in the same consistent way if
substructures should be detected. Combining data from many different
studies may in fact create unphysical structures in abundance space.

An example of a homogeneous treatment of a large number of stars is the
SEGUE survey.\footnote{SEGUE is the Sloan Extension for Galactic
  Understanding and Exploration to map the structure and stellar
  makeup of the Milky Way Galaxy using the 2.5\,m telescope at Apache
  point \citep{SEGUE}.} An interesting illustration of the (inherent?)
difficulties in reaching accurate results is given by the first paper
on the SSPS pipeline used to analyse the SEGUE
spectra \citep{lee2008}.  The pipeline implements eleven methods to derive
iron abundances. Figure~\ref{fig:lee2008} summarizes the resulting
differences between the adopted iron abundance and those derived using
the eleven different methods as a function of signal-to-noise ratios
in the stellar spectra. Most of the methods converge towards the
adopted value around a signal-to-noise ratio of about 25, 
where the typical scatter for any of the methods is about 0.1~dex. 
However, there are methods that give iron abundances that 
deviate systematically by a similar amount also at higher signal-to-noise ratios,
even though in this case the underlying assumptions are quite uniform.
Thus this comparison suggests that the precision (as judged from the
scatter of individual methods) is about 0.1~dex, and that systematic errors 
could be at least as large.


It is possible to access the precision in the derived elemental
abundances with a full forward modelling of the analysis of a stellar
spectrum. This can be done as a preparatory step for instrument
designs. A recent example is given by \citet{2013AN....334..197C} who ran
model spectra through a simulator built to resemble the 4MOST
multi-object spectrograph for VISTA \citep{2012SPIE.8446E..0TD}. The simulator
includes a transmission model of the Earth's atmosphere, a model for
the seeing and sky background, and a simple model of the instrument.
They found that they could reproduce the input abundance ratios with a
precision of 0.1~dex for most elements and 0.2~dex for some
elements.

We note with some interest that Eq.~(\ref{eq:Rmin}) fits results from
recent works in the literature. For example, \citet{nissen2010} used about 100
stars in their study, and $r_{\rm min}$ according to Eq.~(\ref{eq:Rmin}) is
thus 4.4. Since their quoted precision is 0.04~dex for [Mg/Fe], the
difference of about 0.2~dex seen in Fig.~\ref{fig:ill1} is compatible with
the prediction from Sect.~\ref{sect:invest} that the minimum discernible 
difference should be about $r_{\rm min} \times \sigma = 0.17$~dex. 
A similar comparison can be made for the results from
SEGUE, e.g., as reported in \citet{lee2011L}.  With 17\,000 stars
$r_{\rm min}$ is 1.6. The quoted precision is no more than 0.1~dex in
[$\alpha$/Fe] which leads to $r_{\rm min} \times \sigma =
0.16$~dex. Figure~\ref{fig:ill1} shows that the difference between the
thin and the thick disk in the solar neighbourhood may be as large
as 0.2~dex, hence the SEGUE spectra should be able to detect the 
difference between the two disks. \citet{lee2011L} do indeed find
a clearly bimodal distribution in [$\alpha$/Fe], although it may be less
visible once the data are corrected for selection effects \citep{bovy2012}.  
We note that
even if the precision would be somewhat worse the situation is still
good. These two studies nicely illustrate the trade-off between high
precision and high numbers of stars. It also illustrates that our formula
in Eq.~(\ref{eq:Rmin}) is a good representation of actual cases and 
can be used for decision making when planning a large survey or a 
small study.

A differential study is the best way to reach high precision
\citep[e.g.,][]{gustafsson2004,baschek1991,magain1984}. One important
aspect in the differential analysis is that measurement errors or
erroneous theoretical calculation for $\log gf$-values become
irrelevant.  The power of differential analysis has been amply
exemplified over the past decades
\citep[e.g.,][]{edvardsson1993,bensby2004,nissen2010}.  A very recent
example are the studies of solar twins \citep[][who reached precisions
of $<$0.01~dex]{melendez2012}. Such precision is possible because they
study solar twins -- all the stars have very similar stellar
parameters. This means that erroneous treatment of the stellar
photosphere and the radiative transport, as well as erroneous $\log
gf$-values, cancel out to first order. This ``trick'' can be repeated
for any type of star and has, e.g., been successfully applied to
metal-poor dwarf stars
\citep{magain1984,2002A&A...390..235N,nissen2010}.

Most large studies must by necessity mix stars with different stellar
parameters. However, in future large spectroscopic surveys it will be
feasible, both at the survey design stage and in the
interpretation of the data, to select and focus on stars with similar
stellar parameters. Those smaller, but more precise stellar samples
will yield more information on potential substructures in elemental
abundance space than would be the case if all stars were lumped
together in order up the number statistics.

%

\section{Concluding remarks}
\label{sect:concl}

With the advent of Gaia, the exploration of the Milky
Way as a galaxy will take a quantum leap forward. We
will be working in a completely new regime -- that of Galactic
precision astronomy. Gaia is concentrating on providing the best
possible distances and proper motions for a billion objects across the
Milky Way and beyond. For stars brighter than
17th magnitude radial velocities will also be supplied.
However, for fainter stars no radial velocities will be obtained and
thus no complete velocity vector
will be available. No detailed elemental abundances will be
available for any star based on the limited on-board facilities.

The Gaia project has therefore created significant activity also as
concerns ground-based spectroscopic follow-up. A recent outcome of
that is the approval of the Gaia-ESO Survey proposal, which has been given 300 nights on
VLT \citep{gilmore2012}. In Europe several studies are under way for massive
ground based follow-up of Gaia including both low- and high-resolution 
spectra. The designs include
multiplexes of up to 3000 fibres over field-of-views of up to
5~deg$^2$ \citep{2012SPIE.8446E..0TD,2012SPIE.8446E..0SC,balcells}.  
A number of other projects are currently under way and will also 
contribute relevant data to complement Gaia, even though they were not 
always designed with Gaia in mind. Examples include the on-going APOGEE, 
which will observe about 100\,000 giant stars down to $H=12.5$ at high 
resolution in the near-infrared in the Bulge and Milky Way disk \citep{wilson2010},
and LAMOST, which will cover large fractions of the Northern sky and especially
the anti-center direction \citep{cui2012}. Of particular interest to Gaia and to the
European efforts is the GALAH survey, which will use the high-resolution optical 
multi-object HERMES spectrograph at AAT to do a large survey down to $V=14$ 
\citep{heijmans2012}.
The promise of elemental abundances for hundreds of thousands to millions of
stars across all major components of the Galaxy, spread over much
larger distances than ever before, is very exciting. Here we have
investigated which types of sub-structures in
abundance space that could be distinguished with these
observations.  

Clearly the arguments presented in Sect.~\ref{sect:invest} show that
it is mandatory to strive for the best possible precision in the
abundance measurements in order to detect stellar populations that
differ in their elemental abundances from each
other. Equation~(\ref{eq:Nmin}) gives an estimate of the number
of stars needed to detect sub-structures in abundance space when the
precision is known and can be used as a tool for trade-offs
between number statistics and precision when planning large surveys.

\begin{acknowledgements}
  This project was supported by grant Dnr~106:11 from the Swedish
  Space Board (LL) and by grant number 2011-5042 from The Swedish
  Research Council (SF).  This research made use of the SIMBAD
  database, operated at CDS, Strasbourg, France and the SAGA database
  \citep{2008PASJ...60.1159S}.
\end{acknowledgements}

\bibliographystyle{aa}
\bibliography{referenser}

\end{document}